\documentclass[%
 reprint,
 amsmath,amssymb,
 aps,
]{revtex4-2}

\usepackage{graphicx}
\usepackage{dcolumn}
\usepackage{bm}
\usepackage{bbm}
\usepackage{xcolor}
\usepackage{comment}
\usepackage{subcaption}
\usepackage{babel}
\usepackage{amsmath}
\usepackage{amssymb}
\usepackage{graphicx}

\usepackage{placeins}
\usepackage[justification=raggedright,singlelinecheck=false]{caption}

\newcommand{\ve}[1]{\boldsymbol{#1}}
\newcommand{\mat}[1]{\widehat{#1}}

\begin{document}

\title{Superconductivity in a Chern band: effect of time-reversal-symmetry breaking on superconductivity}

\author{Bernhard E.~L\"uscher}
 \affiliation{Department of Physics, University of Zurich, Winterthurerstrasse 190, 8057 Zurich, Switzerland}
\author{Mark H.~Fischer}%
 \affiliation{Department of Physics, University of Zurich, Winterthurerstrasse 190, 8057 Zurich, Switzerland}

\date{\today}

\begin{abstract}
Time-reversal-symmetry breaking is generally understood to be detrimental for superconductivity.
However, recent experiments found superconductivity emerging out of a normal state showing a finite anomalous Hall effect, indicative of time-reversal-symmetry breaking,  in diverse systems from kagome metals, $1T'$-WS$_2$, to twisted MoTe$_2$ and rhombohedral graphene. 
Motivated by these findings, we study the stability of superconducting orders and the mechanisms that suppress superconductivity in the prototypical anomalous Hall system, the Haldane model, where complex hopping parameters result in loop-current order with a compensated flux pattern.
We find that neither spin-singlet nor spin-triplet states are generically suppressed, but the real-space sublattice structure plays a crucial role in the stability of the orders.
Interestingly, the nearest-neighbor chiral states of $d\pm id$ or $p\pm i p$ symmetry couple linearly to the flux, such that the two otherwise degenerate chiralities split under finite flux.
As an experimental probe to distinguish the various orders in this system, we study the anomalous thermal Hall effect, $\kappa_{xy} / T$, which vanishes at zero temperature for topologically trivial superconducting states, but reaches a finite value corresponding to the Chern number in a topologically non-trivial superconducting state.
Our results illustrate that broken time-reversal symmetry through a finite flux is neither generically destructive for superconductivity, nor does it imply non-trivial topological order of the emerging superconducting state. However, in the case of multiple competing pairing channels, the loop-current order can favor a chiral superconducting state.
\end{abstract}

\maketitle

\section{Introduction}
Recently, several material platforms have shown superconductivity emerging out of a normal state exhibiting an anomalous Hall effect (AHE). In particular, such a situation was found in twisted MoTe$_2$~\cite{xu2025signatures}, as well as in rhombohedral graphene~\cite{han2025signatures}.
Furthermore, in the kagome metals of the AV$_3$Sb$_5$ (A=K, Rb, Cs) family~\cite{guguchia2023tunable, mielke2022time, xu2022three, khasanov2022time, feng2021chiral}, superconductivity was found to coexist with an unconventional AHE~\cite{yang:2020} without signs of magnetism~\cite{ortiz:2019a}. Finally, in 1$T'$-WS$_2$ under hydrostatic pressure, an anomalous Hall phase was found, out of which superconductivity emerges~\cite{hossain2025tunable}. While both the rhombohedral graphene and twisted MoTe$_2$ are believed to have a valley- and spin-polarized normal state, in the kagome materials the signatures of time-reversal-symmetry breaking were interpreted as stemming from a loop-current or flux-density-wave order~\cite{jiang2021unconventional,lin2021complex,denner2021analysis,fu2024exotic,zhan2025loop}, which could give rise to non-trivial Chern bands~\cite{wang:2024tmp}.  These findings have renewed attention to the implications of a normal state being a partially-filled topological band structure without time-reversal symmetry on superconductivity.

Superconductivity emerging out of  topologically non-trivial bands  has attracted most attention recently in the context of flat bands, such as multilayer graphene structures~\cite{peotta2015superfluidity}. Concretely, most theoretical studies project the bandstructure to a single band, such that the properties of the band structure are captured by Berry-curvature and quantum-geometry effect. The question then is on the possibly unconventional pairing channels, which might be enforced by the non-trivial bandstructure, leading to topologically non-trivial superconductivity~\cite{murakami:2003, simon:2022, wolf2022topological,wang:2024tmp,wang:2025tmp, maymann:2025tmp,peotta2015superfluidity}.

Finally, signatures of a time-reversal-symmetry-broken normal state in the superconducting phase were studied, such as in the Rashba system with a Zeeman field~\cite{ojanen:2013, sacramento:2014, chung:2014}.
However, less focus was directed on how the fact that time-reversal symmetry (TRS) is broken in a Chern band affects the instability and the resulting ground state.

TRS was early recognized to be a crucial symmetry for superconductivity~\cite{anderson1959theory}. In particular, a finite magnetic field coupling to the electrons' spin though the Zeeman term suppresses spin-singlet superconductivity, such that an unconventional superconducting order is expected in a system with finite magnetic moment~\cite{huxley:2015}. This then raises the question whether in the case of an anomalous Hall system, where the anomalous Hall effect derives from a finite Berry curvature, we generically should expect unconventional pairing, or whether the pairing could equally well be conventional.

To answer these questions, we study superconductivity in the prototypical lattice model featuring an anomalous Hall effect, the Haldane model on the hexagonal lattice~\cite{haldane1988model}. In the Haldane model, the complex next-nearest-neighbor (NNN) hopping amplitudes lead to a (compensated) flux pattern and open a gap at the $K$ and $K'$ points in the Brilluoin zone. In the metallic regime, the anomalous Hall conductivity is non-zero but not quantized. We then start from a tight-binding model on the hexagonal lattice and introduce interactions up to next-nearest neighbors and investigate the stability of the resulting states~\cite{brydon2019loop, pangburn2023superconductivity, awoga2023superconductivity} against introducing TRS breaking in the form of a finite flux. Given the simplicity of the model, we can treat the self-consistency equation using the full band-structure, with no need of projecting to a single-band description.

The TRS breaking of the normal state is generally expected to imprint signatures on the superconducting state. 
As an experimental signature of the breaking of TRS in the superconducting state, we study the temperature-dependence of the anomalous thermal Hall effect. While the anomalous thermal Hall conductivity vanishes for all superconductors for $T\rightarrow0$ due to the vanishing density of states at the Fermi energy, the temperature dependence can indicate nodal gap structure, and a finite slope indicates a topologically non-trivial pairing state.

The rest of the paper is organized as follows. In Sec.~\ref{sec:Haldane}, we introduce the Haldane model, which features Chern bands and an anomalous (quantum) Hall effect. After introducing the mean-field framework for superconductivity and the possible pairing channels on the hexagonal lattice in Sec.~\ref{sec:SC}, we analyze the effect of the finite flux on the transition temperatures of all pairing channels including up to NNN local structures in Sec.~\ref{sec:Results}. In Sec.~\ref{sec:ATHE}, we study the temperature evolution of the anomalous thermal Hall effect as an experimental probe of the underlying time-reversal-symmetry breaking and the topology of the different states before discussing our results and concluding in Sec.~\ref{sec:concl}.

\section{Haldane Model and Normal State}\label{sec:Haldane}
\subsection{Normal State Hamiltonian}
\begin{figure}
\centering
\includegraphics{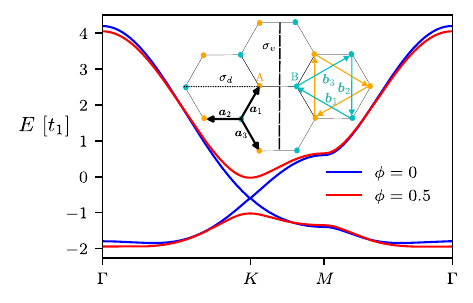}

\caption{Dispersion along high-symmetry lines for the Haldane model with $t_1 = 1$, $t_2 = 0.2$, and $\phi = 0$ and $\phi = 0.5$ respectively. The inset shows the hexagonal lattice with the mirrors $\sigma_v$, $\sigma_d$, and the lattice vectors $\ve{a}_i$ and $\ve{b}_i$.}
\label{fig:HaldaneConventions}
\end{figure}
The Haldane model describes electrons moving on a two-dimensional hexagonal lattice with nearest- and next-nearest-neighbor hopping. The next-nearest-neighbor hopping amplitudes are complex, $t_2 e^{i\phi}$ with $t_2 \in \mathbb{R}$, leading to finite currents on NNN bonds with a corresponding (alternating) flux pattern.  

In the following, we denote by $\ve{a}_i$ the vectors connecting B sites to nearest-neighbor A sites,
\begin{equation}\begin{split}
    \ve{a}_1 &= \frac{1}{2}[1, \sqrt{3}]\\
    \ve{a}_2 &= [-1, 0]\\
    \ve{a}_3 &= \frac{1}{2}[1, -\sqrt{3}],
\end{split}\end{equation}
see inset of Fig.~\ref{fig:HaldaneConventions}, and NNN sites are connected by $\ve{b}_1 = (\ve{a}_2 - \ve{a}_3)$ and cyclic. In momentum space, the Hamiltonian 

reads 
\begin{equation}
    H_0 = \sum_{\ve{k},s} \vec{c}^{\,\dagger}_{\ve{k}s}\mat{H}_0(\ve{k})\vec{c}^{\phantom{\dag}}_{\ve{k}s},
\end{equation}
in the basis of the two-component spinors 
$\vec{c}_{\ve{k}s}^{\,\dagger} = (c_{{\rm A}\ve{k}s}^\dagger, c^{\dagger}_{{\rm B}\ve{k}s})$, where $c^{\dagger}_{a\ve{k}s}$ creates an electron with momentum $\ve{k}$ and spin $s$ on sublattice site $a={\rm A},{\rm B}$, and the conventions we use for the Fourier transform are given in Eq.~(\ref{eq:FourierTransform}). We parametrize the Hamiltonian
\begin{equation}\begin{split}
    \mat{H}_0(\ve{k}) = g_0(\ve{k}) \mat{\tau}^0+\vec{g}(\ve{k})\cdot\vec{\mat{\tau}}\label{eq:H0},
\end{split}\end{equation}
with $\mat{\tau}^{i}$ and $\mat{\tau}^0$ the Pauli matrices and identity acting on the sublattice space, respectively. Further,

\begin{align}
    g_0(\ve{k}) &= 2t_2 \cos(\phi)\sum_n\cos(\ve{k}\cdot\ve{b}_n) -\mu,\label{eq:g0}\\
    g_z(\ve{k}) &= -2t_2 \sin(\phi)\sum_n\sin(\ve{k}\cdot\ve{b}_n ),\label{eq:gz}
\end{align}
with $\mu$ the chemical potential, denote terms connecting the same sublattice, while
\begin{align}
    g_x(\ve{k}) &= t_1 \sum_n \cos(\ve{k}\cdot\ve{a}_n)\label{eq:gx}\\
    g_y(\ve{k}) &= t_1 \sum_n \sin(\ve{k}\cdot\ve{a}_n)\label{eq:gy}
\end{align}
connect different sublattices. 

Figure~\ref{fig:HaldaneConventions} shows the dispersion
\begin{equation}
    \xi_{\ve{k},\pm} = g_0(\ve{k}) \pm |\vec{g}(\ve{k})|
\end{equation}
given by the Hamiltonian defined by Eq.~\eqref{eq:H0} featuring two (spin-degenerate) bands with a gap opening at $K$ and $K'$ for $\phi\neq 0$. 
Note that with both spin components experiencing the same flux, the system explicitly breaks time-reversal symmetry, which manifests itself in an anomalous Hall effect, see below. Finally, we have set the Semenoff (sublattice) mass~\cite{semenoff1984condensed} to zero, as otherwise the combination of mass and $\phi\neq 0$ breaks the $\ve{k} \mapsto -\ve{k}$ symmetry needed for a Cooper instability in any channel.

In the following, we fix the energy scale by setting $t_1=1$ and use $t_2 = 0.2$. We then consider different fillings by changing the chemical potential and study the effect of the phase $\phi$ of the NNN hopping on superconducting pairing instabilities.

\subsection{Anomalous (Quantum) Hall Effect}\label{sec:AHE}
The complex hopping parameters of the Hamiltonian in Eq.~\eqref{eq:H0} lead to a finite anomalous Hall effect as a consequence of the TRS breaking.
The AHE at zero temperature is determined by the integration of the Berry curvature below the Fermi energy, 
\begin{equation}
    \sigma_{xy}=\frac{2 e^2}{\hbar}\frac{1}{N}\sum_{n,\ve{k}}F_{nn}(\ve{k})\Theta(-\xi_{\ve{k},n})
\end{equation}
with $N$ the number of unit cells, $\Theta(.)$ the Heaviside step function, and the factor of two due to spin summation. The Berry curvature is given by 
\begin{equation}\label{eq:Berry}
    F_{nm}(\ve{k})=\partial_xA^y_{nm}(\ve{k})-\partial_yA^x_{nm}(\ve{k}),
\end{equation}
where 
\begin{equation}
    A^{j}_{nm}(\ve{k}) = -i\langle u_{\ve{k},n}|\partial_{k_j}|u_{\ve{k},m}\rangle,
\end{equation}
with $|u_{\ve{k},n}\rangle$ the Bloch vector corresponding to the energy $\xi_{\ve{k},n}$,
\begin{equation}
    \mat{H}_0(\ve{k})|u_{\ve{k},n}\rangle  =  \xi_{\ve{k},n}|u_{\ve{k},n}\rangle.
\end{equation}

\begin{figure}
\centering
\includegraphics[]{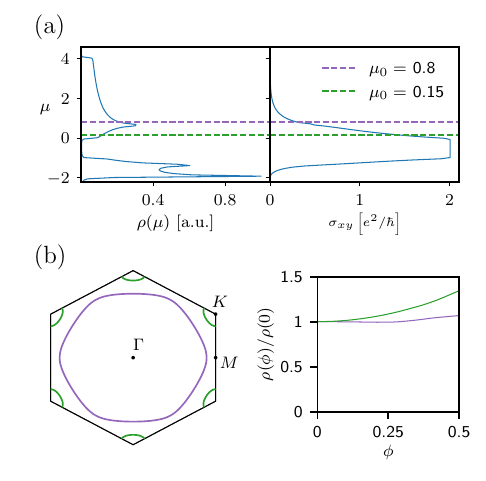}
\caption{\label{fig:Various_Plots_Mus}Haldane model at $\phi = 0.5$. (a) Density of states $\rho(\mu)$ and anomalous Hall conductivity as a function of chemical potential $\mu$. (b) Fermi surfaces for the chemical potentials denoted by dashed lines in (a) usd in Sec.~\ref{sec:Results}. Also shown is the dependence of the density of states on $\phi$ for fixed density of electrons.}
\end{figure}

Figure~\ref{fig:Various_Plots_Mus}(a) shows the anomalous Hall conductivity as a function of the chemical potential $\mu$ for $\phi = 0.5$. Also shown is the density of states 
\begin{equation}
    \rho(\epsilon) = \frac2N\sum_{\ve{k},n}\delta(\epsilon - \xi_{\ve{k},n}).
\end{equation}
For a half-filled system, in other words, a chemical potential within the band gap  and $\phi>0$, $\sigma_{xy}=2 e^2 / \hbar$ is quantized and is proportional to the Chern number of the topological band structure~\cite{thouless:1982}. Consequently, the lower band contributes with a positive Berry curvature, while the upper band contributes a negative Berry curvature to the anomalous Hall conductivity. 

In the following, we focus on two cases: (I) $\mu_0 = 0.8$ and (II) $\mu_0=0.15$ for $\phi = 0$. Case I features a large Fermi surface around the $\Gamma$ point and a rather flat density of states around the chemical potential, but a small $\sigma_{xy}\ll 2 e^2 / \hbar$. Case II, in contrast, features small Fermi pockets around the $K$ and $K'$ points and a stronger energy dependence of the density of states, but a rather large anomalous Hall conductivity. The two cases are shown in Fig.~\ref{fig:Various_Plots_Mus}(b). Note that when changing the flux $\phi$, we fix the density of particles in the system, rather than the chemical potential.

\section{Superconductivity}\label{sec:SC}
\subsection{Pairing interaction and pairing channels}
In order to study the effect of TRS breaking in the form of an anomalous Hall effect (finite flux pattern) on superconductivity, we introduce a pairing interaction
\begin{equation}
    H'=\frac{1}{2}\sum_{\ve{k},\ve{k}'}V^{ab,cd}_{s_1s_2,s_3s_4}(\ve{k},\ve{k}')c^{\dagger}_{a\ve{k}s_1}c^{\dagger}_{b-\ve{k}s_2}c^{\phantom{\dag}}_{c-\ve{k}'s_3}c^{\phantom{\dag}}_{d\ve{k}'s_4}
\end{equation}
omitting the summation over repeated sublattice and spin indices for brevity of notation.

We decouple the interaction terms by introducing the mean-field gap functions
\begin{equation}
    \Delta^{ab}_{s_1s_2}(\ve{k})=\sum_{\ve{k}}V^{ab,cd}_{s_1s_2,s_3s_4}(\ve{k},\ve{k}')\langle c^{\phantom{\dagger}}_{c-\ve{k}s_3}c^{\phantom{\dagger}}_{d\ve{k}s_4}\rangle,
\end{equation}

where $\langle\,. \,\rangle$ denotes a (thermal) expectation value.
In the following, we write the gap functions in terms of the irreducible representations of the point group of the hexagonal lattice ($\phi = 0$). Without spin-orbit coupling, it is sufficient to use the point group $C_{6v}$, which in addition to a six-fold rotation axis possesses three mirrors $\sigma_v$ and three mirrors $\sigma_d$, see inset Fig.~\ref{fig:HaldaneConventions}. For the purpose of defining basis functions, we first introduce the momentum functions 
\begin{equation}\begin{split}
    e_a(\ve{k}) &= \frac{1}{\sqrt{3}}\sum_n \cos(\ve{k} \cdot \ve{a}_n) \\
    e^{\pm}_a(\ve{k}) &= \frac{1}{\sqrt{3}}\sum_n \omega^\pm_n \cos(\ve{k} \cdot \ve{a}_n) \\
    o_a(\ve{k}) &= \frac{1}{\sqrt{3}}\sum_n\sin(\ve{k} \cdot \ve{a}_n)\\
    o^{\pm}_a(\ve{k}) &= \frac{1}{\sqrt{3}}\sum_n\omega^{\pm}_n\sin(\ve{k} \cdot \ve{a}_n),
\end{split}\end{equation}
where $\omega^\pm_n = \exp(\pm i 2\pi n/ 3)$,
and the same functions can be defined for the NNN lattice vectors $\ve{b}_n$~\footnote{Note that with the conventions used in this paper, the basis functions $e^\pm(\ve{k})$ correspond to $d_{x^2-y^2}\mp id_{xy}$-wave order parameters in the continuum limit, while $o^\pm(\ve{k})$ corresponds to $p_x\pm ip_y$.}.
The full gap function can now be written as a direct product of a real-space part $\mat{\Psi}(\ve{k})$, which combines the momentum and sublattice structure, with a spin part $\mat{\zeta}_\Gamma$, 
\begin{equation}
    \mat{\Delta}(\ve{k}) = \sum_{\Gamma} \Delta_{\Gamma} \mat{\Psi}^m_{\Gamma}(\ve{k}) \otimes \mat{\zeta}^m_\Gamma
    \label{eq:factorize_delta}
\end{equation}

with $m$ denoting different basis functions belonging to the same irreducible representation.
The spin structure of the gap function depends on the behavior of the real-space structure under electron exchange, $\mat{\Psi}(\ve{k})= \pm \mat{\Psi}^T(\ve{-k})$, through the Pauli principle and is given by
\begin{equation}
    \mat{\zeta}_\Gamma = \left\{ \begin{array}{ll} i\mat{\sigma}^y & \mat{\Psi}^m_\Gamma(\ve{k})\; {\rm even}\\ \mat{\sigma}^x & \mat{\Psi}^m_\Gamma(\ve{k}) \; {\rm odd} \end{array}\right. \;.
\end{equation}
Note that, without loss of generality, we have chosen the spin-triplet $d$ vector to be along $z$, as the direction of the $d$ vector is not fixed without spin-orbit coupling.

\begin{table}[t]
    \centering
    \caption{\textbf{Intra-sublattice pairing states:} momentum and sublattice basis functions diagonal in sublattice space. The even states with $\mat{\Psi}(\ve{k})= \mat{\Psi}^T(\ve{-k})$ need to be combined with a spin-singlet configuration, while the odd states, $\mat{\Psi}(\ve{k})=-\mat{\Psi}^T(\ve{-k})$, have a spin-triplet configuration.}
    \label{tab:basisfct_intra}
   \begin{tabular}{l|c|c}
 $\Gamma$& even & odd\\
 \hline
  $A_1$ & $e_b(\ve{k})\mat{\tau}^0$ & -\\ 
  $A_2$ & - &  $o_b(\ve{k}) \mat{\tau}^z$ \\ 
  $B_1$ & - & $o_b(\ve{k}) \mat{\tau}^0$ \\ 
  $B_2$ & $e_b(\ve{k})\mat{\tau}^z$ & - \\ 
$E^{\pm}_1$       & $e_b^\pm (\ve{k}) \mat{\tau}^z$ & $o_b^\pm (\ve{k}) \mat{\tau}^0$ \\
$E^{\pm}_{2}$  & $e_b^\pm (\ve{k}) \mat{\tau}^0$ & $o_b^\pm (\ve{k}) \mat{\tau}^z$\\
\end{tabular}    
\end{table}

Table~\ref{tab:basisfct_intra} shows the basis functions for intra-sublattice pairing for both spin-singlet and spin-triplet combinations, while Tab.~\ref{tab:basisfct_inter} shows the basis functions for inter-sublattice pairing. Note that the basis functions for inter-sublattice pairing always combine two Pauli matrices. This structure follows from the real-space structure of the honeycomb lattice, where in any given direction, there is always only one nearest neighbor.

Finally, we can decompose the interaction in the same channels, 
\begin{equation}
    V^{ab,cd}_{s_1s_2,s_3s_4}(\ve{k},\ve{k}') = -\sum_{\Gamma, m}V_{\Gamma}^m [\Psi^m_\Gamma(\ve{k})]_{ab} [\Psi^m_\Gamma(\ve{k})^{\dagger}]_{cd} \Sigma^{\Gamma}_{s_1s_2;s_3s_4},\label{eq:interaction}
\end{equation}
as done in App.~\ref{app:interaction} for on-site (OS), nearest-neighbor (NN), and NNN interactions. Note that in the following, we consider attractive interactions, $V^{m}_{\Gamma}>0$.
The spin structure is again determined by the parity of the basis function $\mat{\Psi}^m_\Gamma(\ve{k})$, with
\begin{equation}
    \Sigma^{\Gamma}_{s_1s_2;s_3s_4} = \left\{ \begin{array}{ll}
    \frac{1}{2} i\sigma^y_{s_1s_2}\left(i\sigma^y\right)^{\dagger}_{s_3s_4} & \mat{\Psi}^m_\Gamma(\ve{k})\;{\rm even}\\
    \frac12 \sum_i (\sigma^i i\sigma^y)_{s_1s_2}(\sigma^i i \sigma^y)^\dagger_{s_3s_4} & \mat{\Psi}^m_\Gamma(\ve{k}) \; {\rm odd}
    \end{array}\right. .
\end{equation}

\begin{table}[t]
    \centering
    \caption{\textbf{Inter-sublattice pairing states:} momentum and sublattice basis functions off-diagonal in sublattice space. The even states with $\mat{\Psi}(\ve{k})= \mat{\Psi}^T(\ve{-k})$ need to be combined with a spin-singlet configuration, while the odd states, $\mat{\Psi}(\ve{k})=-\mat{\Psi}^T(\ve{-k})$, have a spin-triplet configuration.}
    \label{tab:basisfct_inter}
    \begin{tabular}{l|c|c}
 $\Gamma$& even & odd \\
 \hline
  $A_1$ & $\left[e_a(\ve{k}) \mat{\tau}_x + o_a(\ve{k}) \mat{\tau}_y\right]  $ & - \\ 
  $B_2$ & - & $\left[o_a(\ve{k}) \mat{\tau}_x - e_a(\ve{k}) \mat{\tau}_y\right] $  \\
  $E^{\pm}_1$ & - & $\left[o_a^\pm(\ve{k})\mat{\tau}_x - e_a^\pm(\ve{k}) \mat{\tau}_y\right] $  \\
  $E^{\pm}_{2}$  & $\left[e_a^\pm (\ve{k}) \mat{\tau}_x + o_a^\pm(\ve{k})\mat{\tau}_y\right]$ & - \\
\end{tabular}
\end{table}

\subsection{Critical Temperature $T_{\rm c}(\phi)$}

We can use Eqs.~\eqref{eq:factorize_delta} and \eqref{eq:interaction} to write the self-consistency equation for the superconducting gap functions for each interaction channel separately. 

To determine the critical temperature, we consider the linearized gap equation~\cite{mineev1999introduction} and neglect coupling between different order parameters that can arise due to the flux term, to find
\begin{equation}\label{eq:linearized_gap}
    1=V_{\Gamma} T\sum_{\ve{k},\omega_n}\mathrm{Tr}\left[\mat{\Psi}_\Gamma^\dagger(\ve{k})\mat{G}_0(\ve{k},i\omega)\mat{\Psi}^{\phantom{\dagger}}_\Gamma(\ve{k})\mat{G}_0(-\ve{k},-i\omega)^T\right]
\end{equation}
with $\omega_n=(2\pi n + 1)/\beta$ the fermionic Matsubara frequencies and $\beta = 1/T$. Note that the trace in Eq.~(\ref{eq:linearized_gap}) is over orbital indices. We have further introduced the normal state Green's function
\begin{equation}\begin{split}
    \mat{G}_0(\ve{k},i\omega_n) =&\left[i\omega_n-\mat{H}_0(\ve{k})\right]^{-1}\nonumber\\
    =&\left[G_{0+}(\ve{k},i\omega_n)\mat{\tau}_0+G_{0-}(\ve{k},i\omega_n)\hat{g}(\ve{k})\cdot\vec{\mat{\tau}}\right],\label{eq:G0}
\end{split}\end{equation}
with $\hat{g}(\ve{k})=\vec{g}(\ve{k})/|\vec{g}(\ve{k})|$ and
\begin{equation}
    G_{0\pm}(\ve{k},i\omega_n)=\frac{1}{2}\left(\frac{1}{i\omega_n-\xi_{\ve{k},+}} \pm \frac{1}{i\omega_n-\xi_{\ve{k},-}}\right).
\end{equation}

In the following, we study the behavior of $T_{\rm c}$ upon increasing the flux $\phi$ in Eqs.~\eqref{eq:g0} and \eqref{eq:gz}. Before we analyze the different order parameters in detail, we start with a brief discussion of the mechanisms we expect to change the critical temperature.
An important suppression mechanism for superconductivity is  to remove or at least weaken the Cooper logarithm crucial for the pairing instability by leading to pairing of electrons with different energies~\footnote{That that given the matrix structure of the order parameter, intra-band pairing is already not guaranteed without flux~\cite{fischer2013gap, ramires2018tailoring}.}. This effect is most transparent for the suppression via Zeeman coupling, where the energies of the two spin species shift against each other. As we will see in the following, this mechanism also plays out in the situation of loop currents, albeit to a smaller degree.

A second mechanism, however, concerns the density-of-states change and Fermi-surface deformation upon changing the flux. To illustrate this mechanism, we consider a single band crossing the Fermi energy and a spin-singlet superconductor with order parameter $\mat{\Delta}(\ve{k})=\psi(\ve{k})i\mat{\sigma}^y$. The critical temperature can be calculated via an eigenvalue problem~\cite{sigrist2005introduction},
\begin{equation}
    \lambda\psi(\ve{k})=-\rho(0)\langle V(\ve{k},\ve{k}')\psi(\ve{k}')\rangle_{\ve{k}'\,\mathrm{FS}},
\end{equation}
where $V(\ve{k},\ve{k}')$ is the interaction and $\langle\,.\rangle_{\ve{k}'\,\mathrm{FS}}$ denotes a Fermi-surface average. In the weak-coupling limit, the critical temperature $T_{\mathrm{c}}\sim \exp(-1/\lambda)$.
Consequently, the Fermi surface has an influence on $T_{\mathrm{c}}$ via the density of states, as well as its geometry due to the averaging procedure.
Since a change in the flux $\phi$ will also change the Fermi surface in our case, these effects play a role in our modeling. However, this mechanism is rather generic for any change in the dispersion of a system and as such not our main focus. To minimize these effects, the parameters especially of case I are chosen such that no significant change in neither the Fermi surface nor in the density of states takes place over the range of $\phi$ that we considered, as shown in Fig~\ref{fig:Various_Plots_Mus}(b).

\section{Results}\label{sec:Results}
In the following, we present the dependence of the critical temperature of the various pairing symmetries on the flux $\phi$ in the Haldane model. For better comparison between the different pairing channels, we choose  interactions such that the critical temperature for all channels are roughly the same, $T\approx 0.1 t_1$. The interaction parameters are summarized in Table~\ref{tab:PairingInteractions} in Appendix~\ref{app:params}.

\subsection{Intra-sublattice pairing}

We start our discussions with the case of intra-sublattice pairing and, in particular, the trivial sublattice structure, which we write as
\begin{equation}
    \mat{\Delta}(\ve{k}) = \psi_\Gamma(\ve{k}) \mat{\tau}^0.
\end{equation}
Gap functions of this form include the spin-singlet gap functions transforming as $A_1$ and $E_2$, as well as the spin-triplet gap functions with $B_1$ or $E_1$ symmetry, see Tab.~\ref{tab:basisfct_intra}. In this case, the linearized gap Eq.~\eqref{eq:linearized_gap} using the non-interacting Green's function of Eq.~\eqref{eq:G0} yields
\begin{equation}\begin{split}
    1 =  2T  V_\Gamma\sum_{\ve{k},\omega_n}|\psi_\Gamma(\ve{k})|^2 [G_{0+}\tilde{G}_{0+} + G_{0-}\tilde{G}_{0-}\\ - 2 \hat{g}_z(\ve{k})^2 G_{0-}\tilde{G}_{0-}].
    \end{split}
\end{equation}
Here, we have introduced the short forms $G_{0\pm} = G_{0\pm}(\ve{k}, i\omega_n)$ and $\tilde{G}_{0\pm} = G_{0\pm}(-\ve{k}, -i\omega_n)$. After summing over Matsubara frequencies, we find
\begin{equation}\begin{split}
    1 = V_{\Gamma} \sum_{n=\pm}\sum_{\ve{k}} |\psi_\Gamma(\ve{k})|^2 \{[1-\hat{g}_z^2(\ve{k})] S_1(\xi_{\ve{k},n}) \\+ \hat{g}_z^2(\ve{k})S_2(\xi_{\ve{k},n}) \}
    \end{split}
\end{equation}
with
\begin{equation}
    S_1(\xi_{\ve{k},n})=\frac{\tanh(\beta\xi_{\ve{k},n}/2)}{2\xi_{\ve{k},n}},
\end{equation}
and
\begin{equation}
   S_2(\xi_{\ve{k},n})= \frac{\tanh(\beta\xi_{\ve{k},n}/2)}{2 |g_0(\ve{k})|}.
\end{equation}
Importantly, the first term proportional to $S_1(\xi_{\ve{k},n})$ describes intra-band pairing and leads to the Cooper instability at low enough temperature, while the second term describes inter-band pairing. This latter term would only lead to a finite critical temperature above a threshold interaction strength. In general, we thus see that a finite flux indeed suppresses the critical temperature through a finite $\hat{g}_z(\ve{k})$. However, the critical temperature is stabilized by the nearest-neighbor hopping, which enter the denominator of $\hat{g}_z(\ve{k})$. 

\begin{figure}
\includegraphics[]{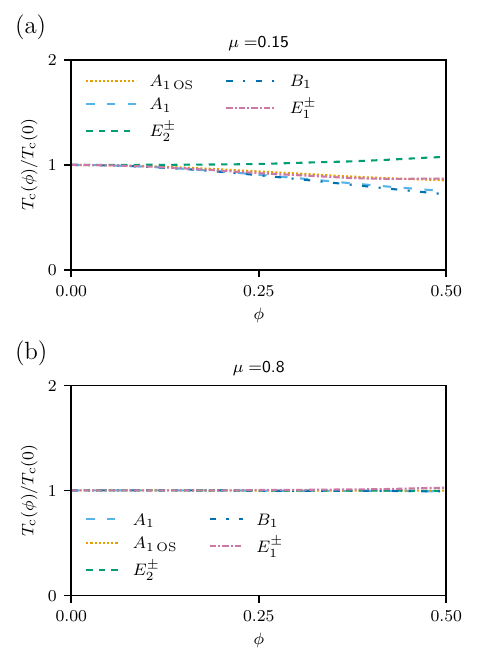}
\caption{\label{fig:TcsIntra} 
Flux dependence of $T_{\mathrm{c}}$ for different intra-sublattice pairing states with the chemical potentials $\mu = 0.15$ (a) and $\mu = 0.8$ (b). For intra-sublattice states, $T_{\rm c}$ is almost independent of $\phi$ for all states.
}
\end{figure}

Next, we discuss the intra-sublattice pairing with opposite sign on the two sublattices,
\begin{equation}
    \mat{\Delta}(\ve{k}) = \psi_\Gamma(\ve{k}) \mat{\tau}^z.
\end{equation}
Gap functions of this form include the spin-singlet gap functions transforming as $B_2$ and $E_1$, as well as the spin-triplet gap functions with $A_2$ or $E_2$ symmetry, see Tab.~\ref{tab:basisfct_intra}. We can again evaluate the linearized self-consistency equation to find
\begin{equation}
    1 = V_{\Gamma} \sum_{\ve{k},n} |\psi_\Gamma(\ve{k})|^2  S_2(\xi_{\ve{k},n}).
\end{equation}
Note that this state never has any intra-band contributions towards Cooper pairing, even in the presence of a finite flux.

Figure~\ref{fig:TcsIntra} shows the critical temperature for intra-sublattice pairing for both spin-singlet and spin-triplet states for on-site (OS) and NNN pairing. Note that we only show the states with a trivial sublattice structure, as the other states ($A_2$ and $B_2$) do not have a Cooper instability. We only find a weak dependence of the states to flux, with a suppression through inter-band pairing that is largely compensated by the (small) increase in the density of states.

\subsection{Inter-sublattice pairing}
\begin{figure}

\includegraphics[]{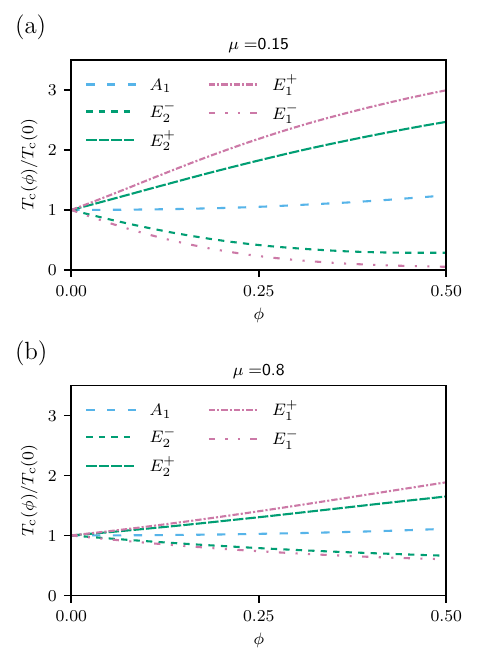}
\caption{\label{fig:TcsInter} 
Flux dependence of $T_{\mathrm{c}}$ for different inter-sublattice pairing states with the chemical potentials $\mu = 0.15$ (a) and $\mu = 0.8$ (b). While the $A_1$ state is almost independent of $\phi$, the chiral states split and show a large dependence.
}
\end{figure}
For the inter-sublattice pairing, we distinguish three cases, namely the case of a spin-singlet gap function with $A_1$ symmetry, which has the same structure as the nearest-neighbor hopping terms, the spin-triplet gap function with $B_2$ symmetry, and, finally, the chiral order parameters of either spin-singlet or spin-triplet structure. In the first case, we find the self-consistency equation
\begin{equation}
    1 = V_{A_1} \sum_{\ve{k},n} \Big[|e_a(\ve{k})|^2 + |o_a(\ve{k})|^2\Big]S_1(\xi_{\ve{k},n}).
\end{equation}
This gap function describes intra-band pairing only and even a finite flux does not change the self-consistency equation. As such, any change in $T_{\rm c}$ stems from band-structure effects such as the shape of the Fermi surface and the density of states.

Next, we discuss the spin-triplet state of $B_2$ symmetry. Here, we find the linearized self-consistency equation
\begin{equation}\begin{split}
    1 = V_{B_2} \sum_{\ve{k},n} \Big[|e_a(\ve{k})|^2 + |o_a(\ve{k})|^2\Big] \Big\{[1-\hat{g}_z^2(\ve{k})] S_2(\xi_{\ve{k},n}) \\+ \hat{g}_z^2(\ve{k})S_1(\xi_{\ve{k},n}) \Big\}.
    \end{split}
\end{equation}
Note that this case is exactly opposite to the situation found for the trivial gap structure above, namely these orders have no Cooper instability for arbitrary small interactions, unless a finite flux is added in the form of the Haldane hoppings with $\phi\neq0$.

Finally, we discuss the inter-sublattice chiral pairing channels, namely the $E_2$ chiral spin-singlet pairing of $d\pm id$ type and the $E_1$ chiral spin-triplet pairing of $p\pm ip$ type, see Table~\ref{tab:basisfct_inter}. For the $E_2$ state, we find

\begin{equation}\label{eq:self_consistency_E2}
\begin{split}
    1 = &V_{E_2} \sum_{\ve{k},n} \Big(|e^{\pm}_a(\ve{k})|^2 \Big\{[1-\hat{g}_y^2(\ve{k})] S_1(\xi_{\ve{k},n})+ \hat{g}_y^2(\ve{k})S_2(\xi_{\ve{k},n}) \Big\}\\
    &  + |o^{\pm}_a(\ve{k})|^2\Big\{[1-\hat{g}_x^2(\ve{k})] S_1(\xi_{\ve{k},n})+ \hat{g}_x^2(\ve{k})S_2(\xi_{\ve{k},n}) \Big\}\\
    & + i\hat{g}_z(\ve{k})[e_a^\pm o_a^{\mp} - e_a^{\mp} o_a^\pm] [S_1(\xi_{\ve{k},+}) - S_1(\xi_{\ve{k},-})]\Big).
\end{split}
\end{equation}

Finally, we note that for the spin-triplet pairing state of $E_1$ symmetry, the linearized self-consistency equation reads

\begin{equation}\label{eq:self_consistency_E1}
\begin{split}
    1 = &V_{E_1} \sum_{\ve{k},n} \Big(|e^{\pm}_a(\ve{k})|^2 \Big\{[1-\hat{g}_x^2(\ve{k})] S_1(\xi_{\ve{k},n})+ \hat{g}_x^2(\ve{k})S_2(\xi_{\ve{k},n}) \Big\}\\
    &  + |o^{\pm}_a(\ve{k})|^2\Big\{[1-\hat{g}_y^2(\ve{k})] S_1(\xi_{\ve{k},n})+ \hat{g}_y^2(\ve{k})S_2(\xi_{\ve{k},n}) \Big\}\\
    & + i\hat{g}_z(\ve{k})[e_a^\pm o_a^\mp - e_a^\mp o_a^\pm] [S_1(\xi_{\ve{k},+}) - S_1(\xi_{\ve{k},-})]\Big).
\end{split}
\end{equation}

Interestingly, the third term in Eqs.~(\ref{eq:self_consistency_E2}) and (\ref{eq:self_consistency_E1}) describes a linear coupling to the flux with opposite sign for the two chiralities, which leads to a splitting of the critical temperature of the solutions with different chirality. Further, the chirality with the higher temperature changes between the two bands. Note that the coupling of chirality to flux is consistent with the findings of Brydon and collaborators~\cite{brydon2019loop}, who 
found that a chiral inter-sublattice pairing on the hexagonal lattice leads to current loops as in the Haldane model. In particular, they identified a bilinear in the order parameter of the form $\mat{\Delta}^{\phantom{\dag}}_{\ve{k}} \mat{\Delta}_{\ve{k}}^\dagger - \mat{\Delta}_{\ve{k}}^\dagger \mat{\Delta}^{\phantom{\dag}}_{\ve{k}}$ which couples to current-loop order. To make the connection more transparent, we write the coupling term as

\begin{equation}
\begin{split}
i\left(e_a^\pm o_a^\mp - e_a^\mp o_a^\pm\right) = \frac{1}{4} \mathrm{Tr}&\left\{\mat{\tau}_z[\mat{\Psi}_{E^{\pm}}(\ve{k})\mat{\Psi}^{\dagger}_{E^{\pm}}(\ve{k})-\right.\\
&\left.\mat{\Psi}^{\dagger}_{E^{\pm}}(\ve{k})\mat{\Psi}_{E^{\pm}}(\ve{k})]\right\},
\end{split}
\end{equation}
where the identity holds for both the $E^{\pm}_1$- and $E^{\pm}_2$-states.

Figure~\ref{fig:TcsInter} shows the dependence of the critical temperature of inter-sublattice pairing states. Indeed, the chiral states belonging to either $E_1$ or $E_2$ split into the different chiralities, while the $A_1$ state is largely uneffected by the time-reversal symmetry breaking in the form of loop currents. Note that here again, we only show pairing channels that have a Cooper instability for arbitrary small interactions.

\subsection{Symmetry considerations}
We finish this section with a short discussion of the symmetry of the system in the presence of the flux term in the Haldane model.
While the hexagonal lattice has point group symmetry $C_{6v}$---we focus only on the two-dimensional lattice symmetries---the fluxes in the Haldane model lower the symmetry to $C_{6v}(C_6)$. Put differently, the mirrors $\sigma_v$ and $\sigma_d$, which are not in $C_6$ have to be combined with time reversal in order to be a symmetry. This symmetry reduction has two consequences: First, the two-dimensional irreducible representations $E_1$ and $E_2$ can in principle be split, as we have seen for nearest-neighbor pairing. In the language of Ginzburg-Landau theory, this is captured by the free energy to quadratic order,
\begin{equation}
    F[\ve{\eta};T] = a(T) |\ve{\eta}|^2 + i \gamma A_{\phi} (\eta_x^* \eta_y - \eta_x \eta_y^*),
\end{equation}
where $a(T) = a_0 (T-T_c)$ with $a_0 > 0$, $\eta_x$ and $\eta_y$ are two (real) basis functions of either $E_1$ or $E_2$, $A_\phi$ is the term representing the finite flux, in other words $A_{\phi=0} = 0 $, and $\gamma$ is a coupling constant. Note that $A_\phi$ transforms as $A_2$ in $C_{6v}$, in other words like a magnetic moment. For $A_\phi \neq 0$, the second term splits the critical temperature for the chiral solutions $\ve{\eta} = \eta_0 (1, \pm i)$.

Second, the finite flux can also couple order parameters that are transforming as different irreducible representations in $C_{6v}$ but belong to the same irrep in $C_6$. These are $A_1$ with $A_2$ and $B_1$ with $B_2$. However, without spin-orbit coupling, the coupling is restricted to be in the same spin-pairing channel. In this second case, the Ginzburg-Landau free energy reads
\begin{equation}
    F[\eta, \eta'; T] = a(T) |\eta|^2 + a' |\eta'|^2 + i \gamma_{\eta\eta'} A_\phi (\eta^* \eta' + c.c.)
\end{equation}
with $\eta$ and $\eta'$ the order parameters belonging to different irreps and $\gamma_{\eta\eta'}$ the respective coupling constant.
As a result, the leading instability ($\eta$), for which $a(T)$ changes sign at $T_{\rm c}$, induces a subleading pairing channel $\eta'$, for which $a' > 0$. Note that the two orders are coupled with a relative phase $\pm\pi/2$, reflecting the time-reversal-symmetry breaking of the normal state. Given the order parameters discussed above for up to NNN pairing, only the NNN spin-triplet pairing state of $B_1$ symmetry couples to the (nearest-neighbor) $B_2$ state. This latter state has no Cooper instability without flux, such that we neglected this coupling in our analysis of the critical temperature.

\section{Anomalous Thermal Hall Conductivity}\label{sec:ATHE}
With superconductivity emerging out of a normal state breaking time-reversal symmetry, we expect signatures of TRS breaking independent of the pairing channel. As such, the system should, among other signatures, exhibit a finite Hall effect and, in the case of a sample breaking inversion symmetry, also a superconducting diode effect~\cite{ando2020observation, daido:2022}, as recently observed for CsV$_3$Sb$_5$~\cite{ge:2025tmp}. In the following, we study another signature of TRS breaking, the thermal Hall conductivity $\kappa_{xy}$, which in addition to indicating TRS breaking also allows us to identify topologically non-trivial superconductivity. Specifically, while the thermal Hall conductivity is finite upon entering the superconducting state and vanishes for all pairing states upon approaching $T\rightarrow 0$, the approach to zero differs dramatically. In particular, $\kappa_{xy}/T$ is integer valued for a topological superconductor, while $\kappa_{xy}/T$ is zero for a trivial superconductor. 

In order to calculate the temperature dependence of $\kappa_{xy}$, we first compute the superconducting gap magnitude $\Delta_\Gamma$ as a function of temperature.
For this purpose, we minimize the free energy~\cite{altland2010condensed,dalal2023field}
\begin{equation}
    F(\Delta_{\Gamma},T) = \frac{N}{V_{\Gamma}}|\Delta_{\Gamma}|^2-T\sum_{\ve{k},n}\log\left(1+e^{-\beta E_{\ve{k},n}}\right),
\end{equation}
where $E_{\ve{k},n}$ are the eigenenergies of the BdG Hamiltonian for a particular pairing channel with $\mat{\Delta}_{\Gamma}(\ve{k})=\Delta_{\Gamma}\mat{\Psi}_{\Gamma}(\ve{k})\otimes\mat{\zeta}_{\Gamma}$~\footnote{Note that the minimization of the free energy leads to the the self-consistency equation for the gap, which provides an alternative route to calculate the temperature dependence of the gap function. However, we have found this to be less stable numerically.
}. 

\begin{figure}
\includegraphics[]{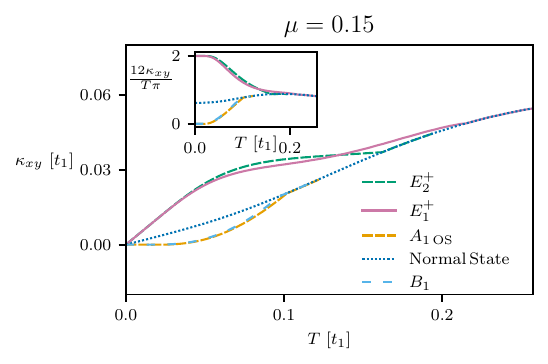}
\caption{\label{fig:Kappas015} Temperature dependence of $\kappa_{xy}$ at $\mu = 0.15$ and $\phi = 0.25$. The inset displays the same quantity but in units such that at $T=0$ it corresponds to the Chern number of the system. Parameters used are summarized in Appendix~\ref{app:params}.}
\end{figure}

The thermal Hall conductivity can be calculated using~\cite{sumiyoshi2013quantum,imai2016thermal}
\begin{equation}\label{eq:kappaxy}
    \kappa_{xy} = -\frac{1}{4\pi T}\int d\epsilon\epsilon^2\Lambda(\epsilon)\frac{\partial f(\epsilon)}{\partial \epsilon},
\end{equation}
where $f(\epsilon)$ denotes the Fermi-distribution function at energy $\epsilon$, and 
\begin{equation}
    \Lambda(\epsilon)=\frac{2\pi}{N}\sum_{\ve{k},n}F_{nn}(\ve{k})\Theta(\epsilon-E_{\ve{k},n})
\end{equation}
is the integrated Berry curvature defined in Eq.~\eqref{eq:Berry} up to energy $\epsilon$. Note that for the superconducting state, $|u_{\ve{k},n}\rangle$ are the eigenvectors of the BdG Hamiltonian.
For low temperatures, the thermal Hall conductivity is linear in temperature, and a Bohr-Sommerfeld expansion yields $\kappa_{xy}\approx\frac{\pi \Lambda(0)}{12}T$. With $\Lambda(0) = N_{\rm C}$, the Chern number of the superconducting state, $12\kappa_{xy}/(\pi T)$ approaches a non-zero integer for a topological state.

\begin{figure}
\centering
\includegraphics[]{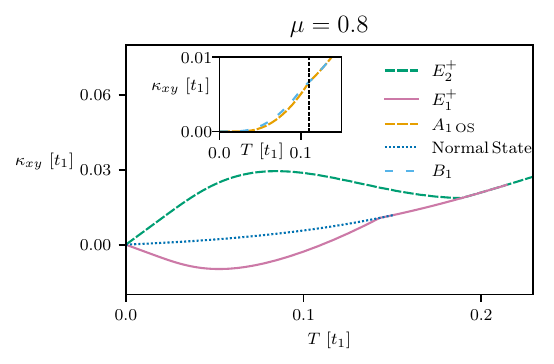}
\caption{\label{fig:Kappas155} Temperature dependence of $\kappa_{xy}$ at $\mu = 0.8$ and $\phi = 0.25$. The inset shows the a zoom-in for the trivial states, with a slightly slower decay for the nodal $B_1$ state. The dashed vertical line indicates the critical temperature of the $A_{1\,\mathrm{OS}}$ state. Parameters used are summarized in Appendix~\ref{app:params}.}
\end{figure}
\subsection{$\mu$ = 0.15}\label{subsec:KappaxyMu015}

Figure \ref{fig:Kappas015} shows the temperature dependence of $\kappa_{xy}$ for various pairing channels for $\mu = 0.15$ with electron pockets around the $K$ and $K'$ points and parameters used as above (see Appendix~\ref{app:params}). For $T\rightarrow0$, the thermal conductivity approaches zero, regardless of the structure of the superconducting order. The topology of the state is encoded in the slope with which it does so. From the inset of Figure \ref{fig:Kappas015} we see, first, that without entering a superconducting state $\frac{12\kappa_{xy}}{T\pi}$ approaches a finite, non-quantized value, in agreement with the finite AHE. Next, for the two chiral states $E_2^{+}$ and $E_1^{+}$ $\kappa_{xy}$ approaches zero with the same slope, indicating identical Chern numbers $N_{\rm C} = 2$ of the two states. Finally, the fast decay of $\kappa_{xy}$ as a function of temperature indicates the topologically trivial nature of both the $A_{1\,\mathrm{OS}}$- and the $B_1$-state. Note that for $\mu=0.15$, both these states are fully gapped. From Fig.~\ref{fig:Kappas015}, we thus see that the topology of the superconducting state is not inherited from the normal-state band structure, but rather depends on the pairing symmetry and its topology. In particular, only the topological superconducting order of $p\pm ip$ ($E_1$) and $d\pm id$ symmetry ($E_2$) lead to a topological state, irrespective of the normal state topology. 

\subsection{$\mu$ = 0.8}\label{subsec:KappaxyMu155}
Figure \ref{fig:Kappas155} shows the temperature dependence of $\kappa_{xy}$ for various pairing channels and $\mu = 0.8$, which features an electron pocket around the $\Gamma$ point, and parameters used as above (see App.~\ref{app:params}). We highlight two differences with respect to the plots for $\mu= 0.15$: First, while the behavior of the $E_2^+$ state is qualitatively the same as for $\mu=0.15$, the behavior of the $E^{+}_1$ state is qualitatively different, in that the Chern number is now  $N_{\rm C} = -1$. 

An intriguing consequence is the sign change of the thermal Hall effect when lowering the temperature.
Second, the small anomalous Hall conductivity of the normal state makes the normal and trivial superconducting states almost indistinguishable. Even when zooming in as done in the inset of Fig.~\ref{fig:Kappas155}, the superconducting transition is barely visible as a small cusp. Nevertheless, we see a slight change in the dependence of the two trivial states upon approach $T=0$, which we associate with the nodal gap structure of the $B_1$ state~\footnote{The fact that this state has point nodes at the Fermi surface means that the Berry field strength is ill-defined. Thus, care has to be taken to not accidentally include a large (but unphysical) contribution in this topologically trivial state.}.

So far, we have only considered the situation, where the chemical potential lies in the upper band. It follows from Eqs.~\eqref{eq:self_consistency_E2} and \eqref{eq:self_consistency_E1} that the opposite chirality is induced for a chemical potential in the lower band. Importantly, this does not change the Chern number of the superconducting states, as the superconducting state now emerges out of hole bands, which have an opposite Chern number for the same chirality as compared to electron bands. As such, the respective behavior on the hole side is almost identical to the electron side.

\section{Discussion and Conclusions}\label{sec:concl}
TRS breaking in the form of loop currents in the Haldane model is not necessary detrimental to superconductivity. More important is the real-space structure of the pairing channels, with some structures, such as the NN $A_1$ state, completely agnostic to the flux in the Haldane model, while others have no Cooper instability in the form of intra-band pairing irrespective of finite loop currents. This dependence of the real-space structure is reminiscent of the situation of spin-singlet superconductivity for an antiferromagnetic spin structure, as discussed by Baltensberger and Straessler~\cite{baltensperger:62}. Importantly, our results demonstrate that time-reversal-symmetry breaking as found in the Haldane model is not intrinsically detrimental for superconductivity, nor does it imply that the dominant superconducting order is topologically non-trivial.

Nevertheless, a non-trivial situation does arise for the case of chiral order parameters. These states' chirality can directly couple to the flux order, resulting in a splitting of the otherwise degenerate transition temperature. Importantly, the real-space structure again plays a crucial role, as only inter-sublattice pairing shows such coupling, a result in agreement with previous results on loop currents induced by chiral superconducting order~\cite{brydon2019loop}.
As such, while the topologically non-trivial bandstructure does not imply non-trivial topology in the superconducting state, the time-reversal-symmetry breaking in the form of loop currents extends the region in parameter space, where a chiral and as such topological superconducting state dominates.

\begin{figure}

\includegraphics[]{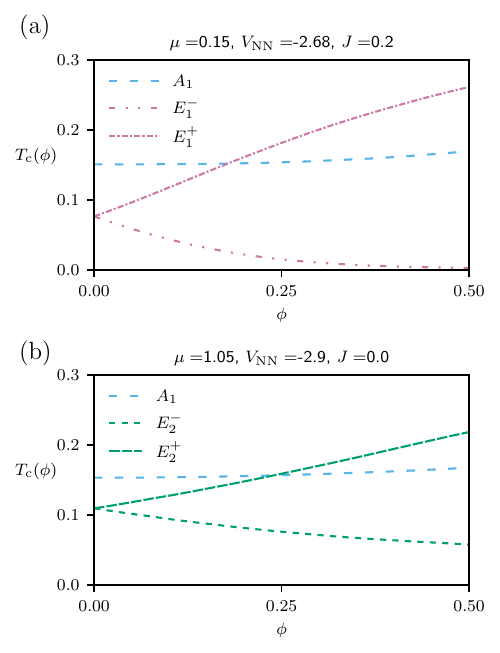}
\caption{\label{fig:TcsInterVJ}
Comparison of the flux dependence of $T_\mathrm{c}$ for the inter-sublattice $A_1$ with $E^{\pm}_1$ and $E^{\pm}_2$ states, respectively, for a given nearest-neighbor density-density ($V_{\mathrm{NN}}$) and spin-spin ($J$) interaction for the two chemical potentials $\mu = 0.15$ (a) and $\mu = 1.05$}
\end{figure}

To illustrate this extended stability region, Fig.~\ref{fig:TcsInterVJ} shows the flux dependence of $T_{\rm c}$ of the spin-singlet $A_1$ state and the two chiral spin-triplet $E^{\pm}_1$ states ($\mu = 0.15$) and the spin-singlet $A_1$ state in combination with the two chiral spin-singlet $E^{\pm}_{2}$ states ($\mu = 1.05$). To compare the instabilities on equal footing, we use the real-space interactions introduced in App.~\ref{app:interaction}. In particular, the chemical potentials and the interaction strengths $V_{\mathrm{NN}}$ and $V_{\mathrm{J}}$ were chosen such that in one case, we are in a parameter range, where the $A_1$ and the $E^{\pm}_1$ states are competitive, and in the other case, the $A_1$ and the $E^{\pm}_2$ states are competitive. Importantly, in both cases one of the chiral states overtakes the $A_1$ state as the dominant instability upon increasing flux.

The effect of the normal-state topology on the leading instability of various tight-binding models on the honeycomb lattice has previously been studied within a weak-coupling renormalization-group approach~\cite{wolf2022topological}. Interestingly, while the dominant superconducting instability depends strongly on the Fermi surface geometry of the normal state, the instability is almost independent of the band topology. 
As we assumed interaction parameters independent of flux, it would be interesting to extend our analysis to the case, where also the effective interactions depend on $\phi$. The TRS breaking in the Haldane model could then enlarge the parameter regime for topological superconductivity.

Unlike previous works on the effect of topologically non-trivial band structures on superconductivity, which project the Hamiltonian to a single band, the simplicity of the Haldane model allows us to treat the self-consistency equations keeping both intra-band and inter-band contributions to pairing in a transparent fashion. As a consequence, the effect of a finite Berry curvature or quantum geometry on the Cooper instability, 
is implicitly taken into account in our analysis. Focusing on the intra-band contributions only in our formulation does, however, still show the effect usually ascribed to the quantum geometry in a formulation with a single band. An example is the dependence on $\hat{g}(\ve{k})$ of the intra-band terms, which is a direct consequence of matrix elements. 

As a physically observable of the different pairing channels and the effect of TRS breaking, we examined the temperature dependence of the anomalous thermal Hall conductivity. While the measurement of this effect is experimentally challenging, especially the $T\rightarrow 0$ behavior to extract the topological nature of the superconducting state, our result indicate that conclusions can be drawn already close below the superconducting transition. In particular, while the topologically trivial states simply decay to zero, the topologically non-trivial state show either an upturn or a sign change upon decreasing temperature.

Finally, we comment on the relevance of our results to more complicated systems like the kagome superconductors of the AV$_3$Sb$_5$ family. While these materials require a more complicated bandstructure involving several orbitals for their low-energy description, we expect the main features we found to translate also to these compounds. In particular, it would be interesting to identify the real-space structures on the kagome lattice that allows for a direct coupling to the proposed flux density-wave order. In addition, a careful study of the thermal Hall effect in this family of superconductors could give important insight into the topological nature of the superconducting ground state.

\acknowledgements
We are grateful for fruitful discussions with Brian M.~Andersen, Matthew Bunney, Titus Mangham-Neupert, Stephan Rachel, Manfred Sigrist, and Ronny Thomale. M.~H.~F. and B.~E.~L. acknowledge support from the Swiss National Science Foundation (SNSF) through Division II (number 207908). B.~E.~L. is further supported by a UZH Candoc Grant (No. FK-24-090).

\appendix
\section{Pairing Interactions for Simulations}\label{app:params}
In Sec.~\ref{sec:Results}, we present the $T_{\mathrm{c}}$'s for different order parameters as a function of $\phi$. For this purpose, we chose for each interaction channel $\Gamma$ an interaction strength $V_{\Gamma}$ such that the $T_{\mathrm{c}}$ at $\phi = 0$ is of the order of $0.1t_1$. We present the interactions for the different parameters in Table~\ref{tab:PairingInteractions}.

\begin{table}[t]
    \centering
    \caption{Strength of pairing interactions used in simulations in the main text.}
    \label{tab:PairingInteractions}
   \begin{tabular}{l|c|c|c}
 $\Gamma$& basis function &$NV_{\Gamma}$ for $\mu = 0.15$&$NV_{\Gamma}$ for $\mu = 0.8$\\
 \hline
 $A_1$&$\mat{\tau}^0$&-1&-0.9\\
  &$e_b(\ve{k})\mat{\tau}^0$&-2.3&-2.6\\ 
  & $e_a(\ve{k}) \mat{\tau}_x + o_a(\ve{k}) \mat{\tau}_y  $ &-1.7&-1.35\\
  $B_1$ &  $o_b(\ve{k}) \mat{\tau}^0$&-1.2&-1.5 \\ 
  $E^{\pm}_1$ &  $o_b^\pm (\ve{k}) \mat{\tau}^0$&-3.1&-2.4\\
  & $o_a^\pm(\ve{k})\mat{\tau}_x - e_a^\pm(\ve{k}) \mat{\tau}_y $ &-1.4&-1.1 \\
$E^{\pm}_{2}$  &$e_b^\pm (\ve{k}) \mat{\tau}^0$ &-1.9&-1.3\\
& $e_a^\pm (\ve{k}) \mat{\tau}_x + o_a^\pm(\ve{k})\mat{\tau}_y$ &-1.35&-1.4 \\
\end{tabular}    
\end{table}

\section{Real-Space Interactions}\label{app:interaction}
In the main text, we expanded the interaction into pairing channels belonging to different irreducible representations to decouple the linearlized self-consistency equation. Here, we explicitly show this expansion starting from real-space interactions up to NNN. This will allow us to study the competition of different pairing channels  as a function of $\phi$ given a set of real-space interaction parameters.

We consider here an onsite Hubbard term, 
\begin{equation}
\begin{split}
    H_{\mathrm{OS}} = \frac{V_{\mathrm{OS}}}{2}\sum_{i,s,a}&c^{\dagger}_{a i s}c^{\dagger}_{a i \bar{s}}c^{\phantom{\dagger}}_{a i\bar{s}}c^{\phantom{\dagger}}_{a i s},
\end{split}
\end{equation}
a nearest-neighbor density-density interaction
\begin{equation}
\begin{split}
    H_{\mathrm{NN}} = V_{\mathrm{NN}}\sum_{\substack{\langle i,j\rangle\\ s,s'}}&c^{\dagger}_{\mathrm{A}i  s}c^{\dagger}_{\mathrm{B}js'}c^{\phantom{\dagger}}_{\mathrm{B}js'}c^{\phantom{\dagger}}_{\mathrm{A}i s},
\end{split}
\end{equation}
and a next-nearest-neighbor density-density interaction
\begin{equation}
\begin{split}
    H_{\mathrm{NNN}} = V_\mathrm{NNN}\sum_{\substack{\langle\! \langle i,j\rangle\! \rangle ,a\\s,s'}}&c^{\dagger}_{ai s}c^{\dagger}_{ajs'}c^{\phantom{\dagger}}_{ajs'}c^{\phantom{\dagger}}_{ai s},
\end{split}
\end{equation}
where $i$ runs over the unit cells, $a$ runs over sublattice indices, $\bar{s}$ is the opposite of $s$, and $\langle i,j\rangle$ and $\langle\! \langle i,j\rangle \!\rangle$ denote nearest- and next-nearest-neighbor sites,  respectively. Finally, we introduce a nearest-neighbor spin-spin interaction
\begin{equation}
    V_{\mathrm{J}}\sum_{\substack{\langle i,j\rangle\\s_1,s_2\\s_3,s_4}}c^{\dagger}_{\mathrm{A}is_1}c^{\dagger}_{\mathrm{B}js_2}c^{\phantom{\dagger}}_{\mathrm{B}js_3}c^{\phantom{\dagger}}_{\mathrm{A}i s_4}\vec{\sigma}_{s_1,s_4}\cdot\vec{\sigma}_{s_2,s_3}.
\end{equation}
Using the conventions 
\begin{equation}\label{eq:FourierTransform}
\begin{split}
    c^{\phantom{\dagger}}_{\mathrm{B} i} =& \frac{1}{\sqrt{N}}\sum_{\ve{k}}e^{i\ve{k}\cdot\ve{r}_i}c^{\phantom{\dagger}}_{\mathrm{ B}\ve{k}},\\
    c^{\phantom{\dagger}}_{\mathrm{A} i} =& \frac{1}{\sqrt{N}}\sum_{\ve{k}}e^{i\ve{k}\cdot\left(\ve{r}_i+\ve{a}_2\right)}c^{\phantom{\dagger}}_{\mathrm{A}\ve{k}},
\end{split}
\end{equation} 
where $N$ is the number of unit cells and $\ve{r}_i$ points to unit cell $i$, the interactions in momentum space are written as
\begin{equation}
\begin{split}
    H_{\mathrm{OS}}=\frac{V_{\mathrm{OS}}}{2N}\sum_{\substack{\ve{k},\ve{k}'\\s}}&c^{\dagger}_{a\ve{k}s}c^{\dagger}_{a-\ve{k}\bar{s}}c^{\phantom{\dagger}}_{a-\ve{k}'\bar{s}}c^{\phantom{\dagger}}_{a\ve{k}'s}\;,
\end{split}
\end{equation}

\begin{equation}
    H_{\mathrm{NN}} = \frac{1}{2}\sum_{\substack{\ve{k},\ve{k}'\\ s,s'}} v^{\rm NN} (\ve{k}-\ve{k}') c^{\dagger}_{\mathrm{A}\ve{k}s}c^{\dagger}_{\mathrm{B}-\ve{k}s'}c^{\phantom{\dagger}}_{\mathrm{B}-\ve{k}'s'}c^{\phantom{\dagger}}_{\mathrm{A}\ve{k}'s}\;,
\end{equation}
\begin{equation}\label{eq:Hnnn}
    H_{\mathrm{NNN}} = \frac{1}{2}\sum_{\substack{\ve{k},\ve{k}'\\a, s,s'}}v^{\rm NNN} (\ve{k}-\ve{k}')c^{\dagger}_{a\ve{k}s}c^{\dagger}_{a-\ve{k}s'}c^{\phantom{\dagger}}_{a-\ve{k}'s'}c^{\phantom{\dagger}}_{a\ve{k}'s}\;,
\end{equation}
and
\begin{equation}
\begin{split}
    H_{\mathrm{J}}=\frac{1}{2}\sum_{\substack{\ve{k},\ve{k}'\\s_1,s_2\\s_3,s_4}}&v^{\rm J} (\ve{k}-\ve{k}') c^{\dagger}_{\mathrm{A}\ve{k}s_1}c^{\dagger}_{\mathrm{B}-\ve{k}s_2}c^{\phantom{\dagger}}_{\mathrm{B}-\ve{k}'s_3}c^{\phantom{\dagger}}_{\mathrm{A}\ve{k}'s_4}\\
    &\times e^{-i\left(\ve{k}-\ve{k}'\right)\cdot\ve{a}_j}\vec{\sigma}_{s_1s_4}\cdot\vec{\sigma}_{s_2s_3}.
\end{split}
\end{equation}
In the above, we have restricted ourselves to the Cooper channel, in other words we have considered only scattering with zero momentum transfer. Furthermore, we have introduced the momentum structure of the interactions of NN and NNN range,
\begin{align}
    v^{\mathrm{NN}}(\ve{k}-\ve{k}')&=\frac{2V_{\mathrm{NN}}}{N}\sum_{\ve{a}_n}e^{-i\left(\ve{k}-\ve{k}'\right)\cdot\ve{a}_n},\\
     v^{\mathrm{J}}(\ve{k}-\ve{k}')&=\frac{2V_{\mathrm{J}}}{N}\sum_{\ve{a}_n}e^{-i\left(\ve{k}-\ve{k}'\right)\cdot\ve{a}_n},
\end{align}
and
\begin{equation}
    v^{\mathrm{NNN}}(\ve{k}-\ve{k}')=4V_{\mathrm{NNN}}\sum_{\ve{b}_n}\cos[(\ve{k}-\ve{k}')\cdot\ve{b}_n].
\end{equation}
The total interaction Hamiltonian is the sum of the four terms introduced above, which, for convenience, we divide into intra- and inter-sublattice interactions, 
\begin{equation}
    H'=H_{\mathrm{intra}} + H_{\mathrm{inter}}
\end{equation}
with 
\begin{equation}
    \begin{split}
    H_{\mathrm{intra}}&=H_{\mathrm{OS}}+H_{\mathrm{NNN}},\\
H_{\mathrm{inter}}&=H_{\mathrm{NN}}+H_{\mathrm{J}}. 
\end{split}
\end{equation}
We express the two constituents $H_{\mathrm{intra}}$ and $H_{\mathrm{inter}}$ of the interaction Hamiltonian $H'$ in a generalized form as

\begin{equation}
\begin{split}
    H_{\mathrm{intra}}&=\frac{1}{2}\sum_{\substack{\ve{k},\ve{k}'}}V^{\mathrm{intra}\,ab;cd}_{s_1s_2;s_3s_4}(\ve{k},\ve{k}')c^{\dagger}_{a\ve{k}s_1}c^{\dagger}_{b-\ve{k}s_2}c^{\phantom{\dagger}}_{c-\ve{k}s_3}c^{\phantom{\dagger}}_{d\ve{k}s_4},\\
    H_{\mathrm{inter}}&=\frac{1}{2}\sum_{\ve{k},\ve{k}'}V^{\mathrm{inter}\,ab;cd}_{s_1s_2;s_3s_4}(\ve{k},\ve{k}')c^{\dagger}_{a\ve{k}s_1}c^{\dagger}_{b-\ve{k}s_2}c^{\phantom{\dagger}}_{c-\ve{k}s_3}c^{\phantom{\dagger}}_{d\ve{k}s_4},
\end{split}
\end{equation}
where from now on, we omit the sums over sublattice and spin indices. For illustration, we treat here explicitly the case of $H_{\mathrm{NNN}}$. We can write Eq.~(\ref{eq:Hnnn}) as
\begin{equation}
\begin{split}
    H_{\mathrm{NNN}}&=\frac{1}{2}\sum_{\ve{k},\ve{k}'}V^{\mathrm{NNN}\,ab;cd}_{s_1s_2;s_3s_4}(\ve{k},\ve{k}')c^{\dagger}_{a\ve{k}s_1}c^{\dagger}_{b-\ve{k}s_2}c^{\phantom{\dagger}}_{c-\ve{k}s_3}c^{\phantom{\dagger}}_{d\ve{k}s_4}
\end{split}
\end{equation}
and factorize $V^{\mathrm{NNN}\,ab;cd}_{s_1s_2;s_3s_4}(\ve{k},\ve{k}')$ into its momentum-, orbital- and spin part as
\begin{equation}
    V^{\mathrm{NNN}\,ab;cd}_{s_1s_2;s_3s_4}(\ve{k},\ve{k}')=v^{\mathrm{NNN}}(\ve{k},\ve{k}')\Omega_{ab;cd}^{\mathrm{NNN}}\Sigma^{\mathrm{NNN}}_{s_1s_2;s_3s_4}.
\end{equation}
For the orbital part, we associate with sublattice $A$($B$) the index 1(2), to write
\begin{equation}
    \Omega_{ab;cd}^{\mathrm{NNN}}
        =\frac{1}{2}\left(\tau_{ab}^0\tau_{cd}^0+\tau_{ab}^z\tau_{cd}^z\right).
\end{equation}
For the spin part, we write
\begin{equation}
\begin{split}
    \Sigma^{\mathrm{NNN}}_{s_1s_2;s_3s_4}&=\delta_{s_1s_4}\delta_{s_2s_3}\\
    &=\frac{1}{2}\left(\sigma_{s_1s_2}^0\sigma_{s_3s_4}^0+\vec{\sigma}_{s_1s_2}\cdot\vec{\sigma}_{s_3s_4}\right).
    \end{split}
\end{equation}
where we made use of the $SU(2)$ completeness relation
\begin{equation}        2\delta_{\alpha\delta}\delta_{\beta\gamma}-\delta_{\alpha\beta}\delta_{\gamma\delta}=\vec{\sigma}_{\alpha\beta}\cdot\vec{\sigma}_{\gamma\delta}.
\end{equation}
Writing the spin part of the interaction in this form makes it straight-forward to decompose it into its singlet and triplet parts $\Sigma^{\mathrm{NNN}\,\rm s}_{s_1s_2;s_3s_4}$ and $\Sigma^{\mathrm{NNN}\,\rm t}_{s_1s_2;s_3s_4}$ respectively. We have
\begin{equation}
    \Sigma^{\mathrm{NNN}\,\rm s}_{s_1s_2;s_3s_4} = \frac{1}{2}\left(\sigma_{s_1s_2}^y\sigma_{s_3s_4}^y\right)\equiv\Sigma^{\rm s}_{s_1s_2;s_3s_4},
\end{equation}
as well as
\begin{equation}
\begin{split}
    \Sigma^{\mathrm{NNN}\,\rm t}_{s_1s_2;s_3s_4} =& \frac{1}{2}\left(\sigma_{s_1s_2}^0\sigma_{s_3s_4}^0+\sigma_{s_1s_2}^x\sigma_{s_3s_4}^x+\sigma_{s_1s_2}^z\sigma_{s_3s_4}^z\right)\\
    \equiv&\Sigma^{\rm t}_{s_1s_2;s_3s_4}.
    \end{split}
\end{equation}
The generalized interaction $V^{\mathrm{NNN}\,ab;cd}_{s_1s_2;s_3s_4}(\ve{k},\ve{k}')$ can thus be subdivided in its singlet and triplet components as
\begin{equation}
\begin{split}
V^{\mathrm{NNN}\,ab;cd}_{s_1s_2;s_3s_4}(\ve{k},\ve{k}')&=\Lambda^{\mathrm{NNN}\,\rm s}_{ab;cd}(\ve{k},\ve{k}')\Sigma^{\mathrm{NNN}\,\rm s}_{s_1s_2;s_3s_4}\\
&+\Lambda^{\mathrm{NNN}\,\rm t}_{ab;cd}(\ve{k},\ve{k}')\Sigma^{\mathrm{NNN}\,\rm t}_{s_1s_2;s_3s_4}
\end{split}
\end{equation}
The quantity ($\Lambda_{ab;cd}^{\mathrm{NNN}\,\rm t}$)$\Lambda_{ab;cd}^{\mathrm{NNN}\,\rm s}$ is obtained by simultaneously (anti-)symmetrizing $v^{\mathrm{NNN}}(\ve{k}-\ve{k}')\Omega^{\mathrm{NNN}}_{ab;cd}$ w.r.t $\ve{k}'\leftrightarrow-\ve{k}'$ and $c\leftrightarrow d$. Since $\Omega^{\mathrm{NNN}}_{ab;cd}$ is symmetric under $c\leftrightarrow d$ we readily obtain

\begin{equation}
    \begin{split}
        \Lambda^{\mathrm{NNN}\,\rm s}_{ab;cd}(\ve{k},\ve{k}')&=\frac{4V_{\mathrm{NNN}}}{N}\sum_{\ve{b}_j}\cos\left(\ve{k}\cdot\ve{b}_j\right)\cos\left(\ve{k}'\cdot\ve{b}_j\right)\Omega^{\mathrm{NNN}}_{ab;cd}\\
        \Lambda^{\mathrm{NNN}\,\rm t}_{ab;cd}(\ve{k},\ve{k}')&=\frac{4V_{\mathrm{NNN}}}{N}\sum_{\ve{b}_j}\sin\left(\ve{k}\cdot\ve{b}_j\right)\sin\left(\ve{k}'\cdot\ve{b}_j\right)\Omega^{\mathrm{NNN}}_{ab;cd}.
    \end{split}
\end{equation}
It is straight-forwardly verified that 
\begin{widetext}
\begin{equation}
    \begin{split}
        \Lambda^{\mathrm{NNN}\,\rm s}_{ab;cd}(\ve{k},\ve{k}')\!=\!\frac{2V_{\mathrm{NNN}}}{N}\!&\left\{\left[\Psi^{\mathrm{Intra}}_{A_1}(\ve{k})\right]_{ab}\left[\Psi^{\mathrm{Intra}}_{A_1}(\ve{k})^{\dagger}\right]_{cd}+\left[\Psi^{\mathrm{Intra},1}_{E^+_2}(\ve{k})\right]_{ab}\left[\Psi^{\mathrm{Intra,1}}_{E^+_2}(\ve{k})^{\dagger}\right]_{cd}+\left[\Psi^{\mathrm{Intra},1}_{E^-_2}(\ve{k})\right]_{ab}\left[\Psi^{\mathrm{Intra,1}}_{E^-_2}(\ve{k})^{\dagger}\right]_{cd}\right.\\
        &\left.+\left[\Psi^{\mathrm{Intra}}_{B_2}(\ve{k})\right]_{ab}\left[\Psi^{\mathrm{Intra}}_{B_2}(\ve{k})^{\dagger}\right]_{cd}+\left[\Psi^{\mathrm{Intra},2}_{E^+_1}(\ve{k})\right]_{ab}\left[\Psi^{\mathrm{Intra,2}}_{E^+_1}(\ve{k})^{\dagger}\right]_{cd}+\left[\Psi^{\mathrm{Intra},2}_{E^-_1}(\ve{k})\right]_{ab}\left[\Psi^{\mathrm{Intra,2}}_{E^-_1}(\ve{k})^{\dagger}\right]_{cd}\right\},\\
    \end{split}
\end{equation}
and
\begin{equation}
    \begin{split}
        \Lambda^{\mathrm{NNN}\,\rm t}_{ab;cd}(\ve{k},\ve{k}')=\frac{2V_{\mathrm{NNN}}}{N}&\left\{\left[\Psi^{\mathrm{Intra}}_{B_1}(\ve{k})\right]_{ab}\left[\Psi^{\mathrm{Intra}}_{B_1}(\ve{k})^{\dagger}\right]_{cd}+\left[\Psi^{\mathrm{Intra},1}_{E^+_1}(\ve{k})\right]_{ab}\left[\Psi^{\mathrm{Intra,1}}_{E^+_1}(\ve{k})^{\dagger}\right]_{cd}+\left[\Psi^{\mathrm{Intra},1}_{E^-_1}(\ve{k})\right]_{ab}\left[\Psi^{\mathrm{Intra,1}}_{E^-_1}(\ve{k})^{\dagger}\right]_{cd}\right.\\
        &\left.+\left[\Psi^{\mathrm{Intra}}_{A_2}(\ve{k})\right]_{ab}\left[\Psi^{\mathrm{Intra}}_{A_2}(\ve{k})^{\dagger}\right]_{cd}+\left[\Psi^{\mathrm{Intra},2}_{E^+_2}(\ve{k})\right]_{ab}\left[\Psi^{\mathrm{Intra,2}}_{E^+_2}(\ve{k})^{\dagger}\right]_{cd}+\left[\Psi^{\mathrm{Intra},2}_{E^-_2}(\ve{k})\right]_{ab}\left[\Psi^{\mathrm{Intra,2}}_{E^-_2}(\ve{k})^{\dagger}\right]_{cd}\right\}.\\
    \end{split}
\end{equation}
\end{widetext}
Here, we denote with $\Psi_{\Gamma}$ the basis functions introduced in the main text, where the superscript ("Inter") "Intra" stands for an irrep that is (off-)diagonal in sublattice space, and in the case of ambiguity, the number 1(2) is added to refer to irreps proportional to $\tau_0$($\tau_z$) in sublattice space.
Similarly, we find 
\begin{equation}
\begin{split}
V^{\mathrm{OS}\,ab;cd}_{s_1s_2;s_3s_4}(\ve{k},\ve{k}')&=\Lambda^{\mathrm{OS}\,\rm s}_{ab;cd}(\ve{k},\ve{k}')\Sigma^{\rm s}_{s_1s_2;s_3s_4},
\end{split}
\end{equation}
with 
\begin{equation}
    \Lambda^{\mathrm{OS}\,\rm s}_{ab;cd}(\ve{k},\ve{k}') = \Lambda^{\mathrm{OS}\,\rm s}_{ab;cd}=\frac{V_{\mathrm{OS}}}{2N}\left(\tau_{ab}^0\tau_{cd}^0+\tau_{ab}^z\tau_{cd}^z\right).
\end{equation}
Combined with the above, we thus have
\begin{equation}
\begin{split}
    V^{\mathrm{intra}\,ab;cd}_{s_1s_2;s_3s_4}(\ve{k},\ve{k}') = &\left[\Lambda^{\mathrm{OS}\,\rm s}_{ab;cd}+\Lambda^{\mathrm{NNN}\,\rm s}_{ab;cd}(\ve{k},\ve{k}')\right]\Sigma^{\rm s}_{s_1s_2;s_3s_4}\\
    +&\Lambda^{\mathrm{NNN}\,\rm t}_{ab;cd}(\ve{k},\ve{k}')\Sigma^{\rm t}_{s_1s_2;s_3s_4}.
\end{split}
\end{equation}
Given the fully factorized form of the interaction, this relates to Eq.~(\ref{eq:interaction}) of the main text. A similar computation can be carried out for $V^{\mathrm{inter}\,ab;cd}_{s_1s_2;s_3s_4}(\ve{k},\ve{k}')$. The two main differences are that the term $H_{J}$ yields a slightly different spin singlet part
\begin{equation}
    \Sigma^{J\,\rm s}_{s_1s_2;s_3s_4} = -\frac{3}{2}\Sigma^{\rm s}_{s_1s_2;s_3s_4},
\end{equation}
and that $\Lambda^{\mathrm{Intra}\,\rm s/\rm t}_{ab;cd}(\ve{k},\ve{k}')$ is off diagonal in sublattice space. We obtain
\begin{widetext}
\begin{equation}
    \begin{split}
        \Lambda^{\mathrm{NN}\,\rm s}_{ab;cd}(\ve{k},\ve{k}')=&\frac{V_{\mathrm{NN}}}{2N}\left\{\left[\Psi^{\mathrm{Inter}}_{A_1}(\ve{k})\right]_{ab}\left[\Psi^{\mathrm{Inter}}_{A_1}(\ve{k})^{\dagger}\right]_{cd}+\left[\Psi^{\mathrm{Inter}}_{E^+_2}(\ve{k})\right]_{ab}\left[\Psi^{\mathrm{Inter}}_{E^+_2}(\ve{k})^{\dagger}\right]_{cd}+\left[\Psi^{\mathrm{Inter}}_{E^-_2}(\ve{k})\right]_{ab}\left[\Psi^{\mathrm{Inter}}_{E^-_2}(\ve{k})^{\dagger}\right]_{cd}\right\}\\
        \Lambda^{\mathrm{J}\,\rm s}_{ab;cd}(\ve{k},\ve{k}')=&\frac{-3V_{\mathrm{J}}}{2N}\left\{\left[\Psi^{\mathrm{Inter}}_{A_1}(\ve{k})\right]_{ab}\left[\Psi^{\mathrm{Inter}}_{A_1}(\ve{k})^{\dagger}\right]_{cd}+\left[\Psi^{\mathrm{Inter}}_{E^+_2}(\ve{k})\right]_{ab}\left[\Psi^{\mathrm{Inter}}_{E^+_2}(\ve{k})^{\dagger}\right]_{cd}+\left[\Psi^{\mathrm{Inter}}_{E^-_2}(\ve{k})\right]_{ab}\left[\Psi^{\mathrm{Inter}}_{E^-_2}(\ve{k})^{\dagger}\right]_{cd}\right\},
    \end{split}
\end{equation}
as well as
\begin{equation}
    \begin{split}
        \Lambda^{\mathrm{NN}\,\rm t}_{ab;cd}(\ve{k},\ve{k}')=&\frac{V_{\mathrm{NN}}}{2N}\left\{\left[\Psi^{\mathrm{Inter}}_{B_2}(\ve{k})\right]_{ab}\left[\Psi^{\mathrm{Inter}}_{B_2}(\ve{k})^{\dagger}\right]_{cd}+\left[\Psi^{\mathrm{Inter}}_{E^+_1}(\ve{k})\right]_{ab}\left[\Psi^{\mathrm{Inter}}_{E^+_2}(\ve{k})^{\dagger}\right]_{cd}+\left[\Psi^{\mathrm{Inter}}_{E^-_1}(\ve{k})\right]_{ab}\left[\Psi^{\mathrm{Inter}}_{E^-_1}(\ve{k})^{\dagger}\right]_{cd}\right\}\\
        \Lambda^{\mathrm{J}\,\rm t}_{ab;cd}(\ve{k},\ve{k}')=&\frac{V_{\mathrm{J}}}{2N}\left\{\left[\Psi^{\mathrm{Inter}}_{B_2}(\ve{k})\right]_{ab}\left[\Psi^{\mathrm{Inter}}_{B_2}(\ve{k})^{\dagger}\right]_{cd}+\left[\Psi^{\mathrm{Inter}}_{E^+_1}(\ve{k})\right]_{ab}\left[\Psi^{\mathrm{Inter}}_{E^+_1}(\ve{k})^{\dagger}\right]_{cd}+\left[\Psi^{\mathrm{Inter}}_{E^-_1}(\ve{k})\right]_{ab}\left[\Psi^{\mathrm{Inter}}_{E^-_1}(\ve{k})^{\dagger}\right]_{cd}\right\}.
    \end{split}
\end{equation}

Finally, we find
\begin{equation}
    V^{\mathrm{inter}\,ab;cd}_{s_1s_2;s_3s_4}(\ve{k},\ve{k}') = \left[\Lambda^{\mathrm{NN}\,\rm s}_{ab;cd}(\ve{k},\ve{k}')+\Lambda^{\mathrm{J}\,\rm s}_{ab;cd}(\ve{k},\ve{k}')\right]\Sigma^{\rm s}_{s_1s_2;s_3s_4}
    +\left[\Lambda^{\mathrm{NN}\,\rm t}_{ab;cd}(\ve{k},\ve{k}')+\Lambda^{\mathrm{J}\,\rm t}_{ab;cd}(\ve{k},\ve{k}')\right]\Sigma^{\rm t}_{s_1s_2;s_3s_4}.
\end{equation}
\end{widetext}

\bibliography{bibliography}
\end{document}